# Does energy efficiency affect ambient PM2.5? The moderating role of energy investment

Cunyi Yang[1], Tinghui Li[1*], Khaldoon Albitar[2]

[1]School of Economics and Statistics, Guangzhou University, Guangzhou, China
[2]Faculty of Business and Law, University of Portsmouth, Portsmouth, UK

**Abstract**

The difficulty of balance between environment and energy consumption makes countries and enterprises face a dilemma, and improving energy efficiency has become one of the ways to solve this dilemma. Based on data of 158 countries from 1980 to 2018, the dynamic TFP of different countries is calculated by means of the Super-SBM-GML model. The TFP is decomposed into indexes of EC (Technical Efficiency Change), TC (Technological Change) and EC has been extended to PEC (Pure Efficiency Change) and SEC (Scale Efficiency Change). Then the fixed effect model and fixed effect panel quantile model are used to analyze the moderating effect and exogenous effect of energy efficiency on PM2.5 concentration on the basis of verifying that energy efficiency can reduce PM2.5 concentration. We conclude, first, the global energy efficiency has been continuously improved during the sample period, and both of technological progress and technical efficiency have been improved. Second, the impact of energy efficiency on PM2.5 is heterogeneous which is reflected in the various elements of energy efficiency decomposition. The increase of energy efficiency can inhibit PM2.5 concentration and the inhibition effect mainly comes from TC and PEC but SEC promotes PM2.5 emission. Third, energy investment plays a moderating role in the environmental protection effect of energy efficiency. Fourth, the impact of energy efficiency on PM2.5 concentration is heterogeneous in terms of national attribute, which is embodied in the differences of national development, science & technology development level, new energy utilization ratio and the role of international energy trade.

## 1      Introduction

As one of the most serious air pollutants in the world, PM2.5 poses a great threat to the environment and public health, and the use of energy is the main cause of PM2.5 emission. Previous studies have proved that there is a significant correlation between fine particulate pollutants and respiratory morbidity and mortality (Brunekreef and Holgate, 2002). The increase of PM2.5 concentration in the air may directly lead to the increase of morbidity and mortality in the population (Helfand et al., 2001;Nemery et al., 2001). This is because the diameter of PM2.5 is small enough to reach the end of the respiratory tract through filtering nasal hair. With a large surface area, it can carry a variety of toxic substances, which can then be exchanged through the lungs to damage other parts of the body (Xing et al., 2016). As the source of pollutant emission, energy has always been a research hotspot. Developing countries are in the process of urbanization and industrialization, which inevitably consume a lot of fossil energy, so they will also face the huge challenge of PM2.5 pollution (Sati and Mohan, 2014;Wang et al., 2016a;Wang et al., 2016b;Gorelick and Walmsley, 2020). In India, 52% of the annual population-weighted PM2.5 concentration comes from residential energy use emissions and causes 511,000 premature deaths per year (Conibear et al., 2018). China's rapid development increases the consumption of fossil energy in the industrial sector, which becomes the main source of PM2.5 (Xu and Lin, 2018). According to The Global Burden of Disease Study, PM2.5 pollution caused 1.1 million deaths in China in 2015 (Cohen et al., 2017). Southeast Asia experienced severe haze in June 2013 with PM2.5 concentration as high as 329μg/m$^3$(Betha et al., 2014). At present, the developed countries have a good performance in pollution emission, but they cannot be spared (Kemfert and Schmalz, 2019). During 1990-2000, the consumption of coal in Japan increased, and the emissions from the energy sector rose significantly (Sugiyama et al., 2009). In recent decades, air quality regulations in the United States have reduced pollution emissions from traditional fossil energy sources. From 2000 to 2015, the U.S. average annual PM2.5 concentration has decreased by 42%, and the corresponding PM2.5 mortality burden has also decreased significantly (Fann et al., 2017;Zhang et al., 2018). However, recent epidemiological studies have shown that current PM2.5 exposure still has a significant adverse impact on the health of the American people (Bennett et al., 2019;Pope et al., 2019). Therefore, it is of great significance to study the relationship between energy and PM2.5.

Energy efficiency, with complex composition, sources and abundant measurement methods, is the central goal of energy utilization and development. Färe is the first who published a paper on energy efficiency in the field of power plant energy (Färe et al., 1983). Index decomposition analysis (IDA), as a perfect tool for energy policy research and analysis, is one of the main energy research methods (Ang and Zhang, 2000;Ang, 2004). In 2006, Ang proposed an analysis framework based on IDA, which can track the energy efficiency trend of the whole economy (Ang, 2006); Countries such as the United States, Canada and New Zealand have adopted IDA based analytical framework to track the trend of energy efficiency changes within their economies (Ang et al., 2010). In addition to IDA, Data envelopment analysis (DEA) is also an important method for evaluating energy efficiency. At present, many developed countries have adopted the DEA method to evaluate their energy efficiency, such as Canada (Hailu and Veeman, 2001), the United States (Mukherjee, 2008), APEC countries (Hu and Kao, 2007), Japan and OECD countries (Zhou et al., 2007;Honma and Hu, 2008), as well as Asian developing countries such as South Korea and India (Lee et al., 2002;Mukherjee, 2010). As the second largest energy consumer, China has abundant related studies. For example, Zhang et al. used the window DEA, which can measure the efficiency of cross-section and time-varying data, to discuss the total factor energy efficiency and change trend of developing countries (Zhang et al., 2011); Wu et al. employed the DEA model containing $CO_2$ emissions to evaluate the energy efficiency of China's industry, concluding that the energy efficiency of China's industry has increased by 5.6% every year since 1997, and this progress mainly comes from technological progress (Wu et al., 2012). From 2006 to 2015, at least 144 academic papers on energy efficiency measured with the DEA model are published in high-level journals. When the production function between input and output is almost non-existent or extremely difficult to obtain, DEA becomes a good tool for analyzing energy efficiency issues (Mardani et al., 2017).

Changes in energy efficiency often have social and environmental consequences and are themselves influenced by a number of factors. Energy efficiency is generally regarded as the cornerstone of mitigating climate change and achieving sustainable development. Its supporters believe that it plays an outstanding role in energy conservation, environmental improvement, energy security, reducing energy costs, improving economic competitiveness and creating employment opportunities (Schnapp, 2012). According to World Energy Outlook (2012) of International Energy Agency, energy efficiency is an important policy strategy to solve energy security, promote economic progress and reduce greenhouse gas emissions. Perez Lombard et al. believe that many concepts related to energy efficiency, such as efficiency, effectiveness, savings, intensity, performance, etc., are often improperly used by scholars, thus hindering the analysis of energy efficiency. Improving energy efficiency without saving energy does not really solve the global energy challenge (Perez-Lombard et al., 2013). In recent years, studies on factors affecting energy efficiency have also emerged continuously. Backlund et al. believe that the improvement of energy efficiency should combine the investment in energy saving technology with the continuous energy management practice. Integrating the energy management into the future energy policies will play an important role for the energy saving goal (Backlund et al., 2012). Zhang et al. believe that energy policy plays a crucial role in improving energy efficiency (Zhang et al., 2011). In addition, the improvement of green technology and institutional quality will promote energy efficiency, while trade openness and urbanization will reduce energy efficiency (Stern, 2012;Jiang et al., 2017;Lv et al., 2017;Danquah, 2018;Sun et al., 2019). Sun et al. made a comparative study of developed and developing countries, and found that the energy efficiency estimates of the two types of economies seemed to be consistent, except Singapore, Iceland, Algeria, Ghana, Thailand and Iran, which seemed to have a gap with other countries, while the other economies were at the peak of efficiency (Sun et al., 2019). This has also been confirmed to varying degrees in other scholars' studies (Filippini and Hunt, 2011;Stern, 2012;Adom et al., 2018). Although related research is countless, there are two problems exist. First, in the past, the research subjects are usually few (dozens of countries, or even only one economy), and the time span is usually not long, making the analysis of impact factors and heterogeneity lack of sufficient samples and comparison, and the credibility of the conclusions needs to be improved. Second, some studies believed that almost all countries had the highest energy efficiency, while others found that there was a great disparity in energy efficiency among different economies. These situations are due to the distortion of the static comparison of energy efficiency. If we analyze the dynamic change rate of energy, this problem can be alleviated.

The main work and marginal contribution of this paper are as follows: First, the temporal and spatial evolution characteristics of energy TFP are investigated from two dimensions of time and space. Based on the annual energy related input-output indicators of 158 countries from 1980 to 2018, the dynamic energy TFP and the decomposition indicators EC, TC, PEC and SEC are calculated considering $CO_2$ as undesirable output. By analyzing the overall TFP value, it is found that the global energy efficiency and its decomposition indicators are in a sustained growth trend during the sample period. Second, the impact of energy efficiency changes on PM2.5 concentrations is studied. Through fixed effect regression and fixed effect panel quantile regression, it is found that dynamic energy efficiency is negatively correlated with PM2.5 concentration, and the inhibition effect of energy efficiency on PM2.5 concentration is verified by static energy efficiency. Besides, effects of energy efficiency on PM2.5 are heterogeneous among the decomposition indicators, that is, the inhibition effect mainly comes from TC and PEC, while SEC can promote PM2.5 emissions. Third, the role of energy investment in the impact of energy efficiency on PM2.5 is analyzed. It is found that energy investment plays a significant moderating role in this process. Fourth, the heterogeneous impact of energy efficiency on PM2.5 in terms of national attributes is studies. It is found that the environmental effects of energy efficiency are indeed heterogeneous in the four dimensions of national development degree, scientific & technological development level, the proportion of new energy utilization and the role of international energy trade, that is, the environmental protection effect of energy efficiency is greater in developed countries, countries with high levels of scientific & technological development, energy exporting countries as well as countries with low proportion of new energy utilization. In combination, in the dimension of national development degree, technical efficiency change (EC) and technological change (TC) jointly play a decisive role; in the dimension of the level of scientific & technological development, technological change (TC) and pure efficiency change (PEC) play a decisive role; in the dimension of new energy utilization degree, technical efficiency change (EC) plays a decisive role; in the dimension of energy importing and exporting countries, the efficiency change (PEC) plays a decisive role.

The rest of this paper is structured as follows: In the second section, based on the analysis of the relationship between energy efficiency, PM2.5 concentration and energy investment as well as the characteristics of each economy, the research hypothesis of this paper is proposed, and the empirical research model and data sources are introduced. In the third section, the DEA method based on Super-SBM-GML is constructed to measure dynamic energy TFP, and the efficiency value obtained and its decomposition indicators are analyzed, and then the variables are classified and descriptive statistics are carried out. In the fourth section, the influence of energy efficiency and its decomposition indicators on PM2.5 concentration is analyzed, and the moderating role of energy investment is discussed. The fifth section is the additional analysis. Starting from the characteristics of the sample countries, the heterogeneity of the impact of energy efficiency on PM2.5 concentration is analyzed, and the differences of four dimensions including national development level, scientific & technological development level, the proportion of new energy utilization and the role of international energy trade are respectively considered. The sixth part draws the basic conclusion.

## 2    Research design

## 2.1    Research hypotheses

Energy strategy is an important pillar of social development, and sustainable development is also an important issue of global concern. Therefore, policy making and corporate strategy planning of various countries attach great importance to the improvement of energy efficiency (Li et al., 2021a). As the world's largest energy consumer, China's Energy Technology Revolution and Innovation Action Plan (2016-2030) proposes that by 2030, a sound energy technology innovation system suited to China's national conditions will be established, and the overall energy technology level will reach the international advanced level, so as to support the coordinated and sustainable development of the energy industry and the ecological environment. The goal of becoming a world power in energy technology is also put forward. Previous studies have also proved that

China's energy efficiency is in a trend of steady rise (Wu et al., 2012). In the studies on other major energy economies, such as APEC countries, OECD countries, the United States, Japan, Canada and India, the conclusion of energy efficiency growth has also been basically obtained (Hailu and Veeman, 2001;Lee et al., 2002;Hu and Kao, 2007;Zhou et al., 2007;Honma and Hu, 2008;Mukherjee, 2008;2010;Sukharev, 2020). Globally, the existence of technology spillover effect has been widely recognized. With the internalization of foreign direct investment of multinational corporations, energy technology has realized technology transfer. Therefore, global energy technology should be in the process of continuous progress. In addition, production activities are closely related to energy consumption, and the improvement of energy efficiency is also reflected in the improvement of input-output efficiency of global production activities as well as the expansion of production scale effect. Based on this, this paper puts forward the following hypothesis:

**H1:** Global energy efficiency continues to grow during the sample period, reflected in both technological progress and improvement of technical efficiency.

Air pollution, which mainly comes from the emission of fossil energy consumption, has a serious impact on the environment and public health. The current energy technology focuses on the field of environmental protection, and all energy related parties are forced to continuously improve the technology under the responsibility of environmental protection, such as harmless coal mining technology, clean and efficient coal utilization technology, spent fuel reprocessing and safe disposal technology of high-level radioactive waste, energy saving and energy efficiency improvement technology, etc. In addition, with the passage of time, the improvement of energy management efficiency and the accumulation of production experience in production activities will also bring the progress of comprehensive energy efficiency, thus reducing the emission of air pollutants. Based on this, this paper puts forward the following hypothesis：

**H2:** The increase of energy efficiency can inhibit PM2.5 concentration.

There are different opinions on the role of investment in pollution emission. Muhammad et al. found that public private partnerships investment in energy damaged environmental quality by increasing carbon emissions (Shahbaz et al., 2020); Jungho holds that FDI tends to increase carbon dioxide emissions, which proves the pollution haven hypothesis (Baek, 2016); Alex et al. believe that renewable energy and FDI reduce carbon emissions (Acheampong et al., 2019). Karim et al., (2021) concluded that capital expenditure increases carbon emission unless it is a green investment. This paper holds that in the process of PM2.5 concentration inhibited by the improvement of energy efficiency, energy investment plays a moderating role in promoting this inhibition. Based on this, this paper puts forward the following hypothesis:

**H3:** Energy investment plays a moderating role in promoting the inhibition of PM2.5 by energy efficiency.

Different countries are in different dimensions of development degree, scientific & technological level, new energy level and other aspects(Mo et al., 2018;Jiang et al., 2020). These differences lead to variation in energy consumption level, equipment and efficiency level, which will make the environmental protection effect of energy efficiency vary to a certain extent. In international trade, energy exporting countries and energy importing countries are in the state of oversupply and short-supply respectively. There are differences in the actual cost of energy and policy priorities, so the role of energy efficiency is inevitably different. Based on this, this paper puts forward the following hypothesis:

**H4:** The environmental effects of energy efficiency are heterogeneous in the dimensions of national development level, scientific & technological development level, proportion of new energy utilization and role of international energy trade.

## 2.2 Model construction

According to the basic hypotheses proposed in this paper, a fixed-effect panel regression model is established to verify the impact of dynamic energy efficiency on PM2.5 (Li et al., 2020a;Li et al., 2020b;Matei, 2020). The basic form of regression model is as follows:

$$AP_{it} = \alpha_0 + \alpha_1 * TFP_{it} + \alpha_2 * X_{it} + \eta_i + \varepsilon_{it} \quad \#(1)$$
$$AP_{it} = \beta_0 + \beta_1 * EC_{it} + \beta_2 * X_{it} + \eta_i + \varepsilon_{it} \quad \#(2)$$
$$AP_{it} = \gamma_0 + \gamma_1 * TC_{it} + \gamma_2 * X_{it} + \eta_i + \varepsilon_{it} \quad \#(3)$$
$$AP_{it} = \delta_0 + \delta_1 * PEC_{it} + \delta_2 * X_{it} + \eta_i + \varepsilon_{it} \quad \#(4)$$
$$AP_{it} = \theta_0 + \theta_1 * SEC_{it} + \theta_2 * X_{it} + \eta_i + \varepsilon_{it} \quad \#(5)$$

where $i$ represents the individual country; t is the time; *AP* (air pollution) is the explained variable i.e. PM2.5 concentration; *TFP*, *EC*, *TC*, *PEC* and *SEC* are explanatory variables, representing the dynamic total factor productivity of energy, energy technology efficiency change, energy technology change, pure technology efficiency change and scale efficiency change respectively; *X* is the covariate that may affect the degree of air pollution in a country, including the logarithm of net FDI inflow of country $i$ in period $t$, population density and the proportion of current industrial added value in GDP; $\eta_i$ represents individual fixed effect; $\varepsilon_{it}$ is the error term.

The above methods mainly investigate the influence of explanatory variables on the conditional expectation of the explained variables, which is actually regression of the mean value. However, this paper also pays attention to the impact of energy efficiency on the entire conditional distribution. Besides, when using OLS mean regression, the objective function of minimization is the sum of squares of residuals ($\sum_{i=1}^{n} e_i^2$), so it is easily affected by outliers. Koenker and Bassett propose that "quantile regression" not only provides comprehensive information of conditional distribution, but also uses the weighted average of the absolute value of residuals (such as $\sum_{i=1}^{n} |e_i|$) as the objective function of minimization, so it is not easy to be affected by outliers (Koenker and Bassett, 1978). Then, Koenker proposed a fixed-effect panel quantile regression with moderating effect (Koenker, 2004). Previous quantile studies have generally assumed that a single effect only causes a parallel (positional) shift in the distribution of the response variable instead of the entire distribution (Koenker, 2004;Canay, 2011). In other words, it does not take into account the individual heterogeneity that has not been observed. In order to solve this problem, this paper adopts the new MM-QR method to investigate the impact of energy efficiency changes on PM2.5 (Machado and Silva, 2019). MM-QR technology has advantages in estimating panel data models with individual effects and eliminating endogenous problems (Lee et al., 2021). Unlike previous studies, we distinctively apply a quantile MM-QR model to examine the dynamic energy efficiency- PM2.5 emissions nexus. This helps to analyze the impact of the change of dynamic energy efficiency and its decomposition indicators on PM2.5 and can also help in understanding the impact of dynamic energy efficiency on PM2.5 emissions from the perspective of pollution level. This offers a more inclusive understanding of the impact of dynamic energy efficiency on PM2.5 emissions by considering the differences of samples with different pollution levels.

Considering the moderating effect of energy investment in the impact of dynamic energy

efficiency on PM 2.5, the following model is constructed:

$$AP_{it} = \alpha_0 + \alpha_1 * TFP_{it} + \alpha_2 * LnEI_{it} + \alpha_3 * TFP_{it} * LnEI_{it} + \alpha_4 * X_{it} + \eta_i + \varepsilon_{it} \#(6)$$

where LnEI represents the logarithm of energy investment. Similarly, fixed-effect regression model and MM-QR fixed-effect panel data quantile model are adopted in the analysis.

## 2.3 Data sources

In this paper, 158 countries with complete data are selected as the initial research samples. First, when measuring the explanatory variable dynamic energy TFP, a DEA model with global reference set is used, so a full sample base must be built (Li et al., 2019;Li and Li, 2020;Li and Liao, 2020;Zhong and Li, 2020). Data of the input-output indicators are derived from PWT 10.0 (Feenstra et al., 2015), U.S. Energy Information (EIA) and World Bank Database, containing 5875 samples from 158 countries or regions for 1980-2018.

Second, data of the explained variable, PM2.5 concentration, are from World Bank Database. PM2.5 concentration (μg/m$^3$) refers to the average level of a country's population exposed to aerosols with an aerodynamic diameter of less than 2.5μm, which can penetrate the respiratory tract and cause serious health damage. Exposure is calculated by the annual average density of population-weighted PM2.5 in urban and rural areas. Sample years are 1990, 1995, 2000, 2005 and 2010-2017, and the sample size is 1785.

Finally, energy investment is adopted as the moderating variable. As an important carrier of the commercial and financial attributes of energy in the global energy market, listed energy companies play an important role in the energy market and energy investment, and the shareholding of listed energy companies is an important way of energy investment. Data are from global famous listed company database OSIRIS of BVD inc. (https://osiris.bvdinfo.com). The sample companies are selected according to the three types of enterprises in the US SIC code: 12, 13 and 29, representing coal mining industry, oil and gas extraction industry, petroleum refining and related industries respectively. The total assets of the sample companies from 1991 to the end of 2018 are extracted and the countries to which they belong are distinguished. It should be noted that for the purpose of tax avoidance, many enterprises choose foreign countries as Legal addresses, which is contrary to the research purpose of this paper, therefore, the trading address is taken as the actual country where the company belongs, and 3811 companies from 103 countries are finally sampled.

The control variables are net foreign direct investment, population density and the proportion of industrial added value in GDP.

## 3 Measurement of independent variables and descriptive statistics

## 3.1 Measurement and analysis of dynamic energy TFP

In this paper, the data envelope analysis (DEA) method based on the super-efficiency SBM-GML model is used to measure the total factor productivity of national energy. The super-efficiency SBM is a distance function, also known as the "Super Slacks-based Measure Directional Distance", and GML is a panel data model. Data envelopment analysis (DEA) was proposed in 1978 to evaluate the relative efficiency of a group of decision-making units (DMU) with multiple inputs and outputs (A. et al., 1978). The distance functions of the basic model are CCR and BCC models, but they do not take "slack" into account. To make up for this shortcoming, Tone put forward

SBM model and super-efficiency SBM model in 2001 and 2002. The latter not only takes slack variables into consideration, but also can rank decision making units whose efficiency value is greater than 1 (Tone, 2001;2002). Malmquist-TFP index is introduced by Malmquist first and developed in Caves innovative research. It is used to measure TFP changes between two periods. The directional distance function containing undesirable outputs is introduced into the Malmquist index to support the analysis of the undesirable outputs (Malmquist, 1953;Caves, 1982). To facilitate intertemporal comparison and overcome the problem of no feasible solution, Oh included the production unit in the global reference set and constructed the Global-Malmquist-Luenberger (GML) index (Oh, 2010).

The general DEA method only uses the input-output ratio to measure the static efficiency of DMU, while the Malmquist index model based on DEA can analyze the dynamic efficiency variation trend of DMU according to panel data. Decomposition analysis using DEA-Malmquist index model is beneficial to provide more robust and in-depth support for TFP estimation. Malmquist productivity index can be decomposed into the product of EC (Technical Efficiency Change) and TC (Technological Change) (Färe et al., 1992), where EC can be extended to the product of Pure Efficiency Change index (PEC) and Scale Efficiency Change index (SEC) (Färe et al., 1994), as shown in Equation (7).

$$TFP = EC(CRS) * TC(CRS) = PEC(VRS) * SEC(CRS, VRS) * TC(CRS) \#(7)$$

where *TFP* index is dynamic energy TFP index, reflecting the change degree of energy TFP in different countries. *CRS* refers to constant return to scale, and *VRS* refers to variable return to scale. *EC* refers to the deviation between the actual output and the "best practice" output under the given production function. The smaller the deviation is, the higher the efficiency is. It specifically measures the resource allocation level of energy input, the improvement of energy management efficiency and the accumulation of production experience. *TC* represents the frontier movement of production function under the given input of production factors. It measures the invention of energy technology, the fundamental innovation of energy production and operation system and the improvement of technology. Among the two indicators derived from *EC*, *PEC* measures the change proportion of pure technical inefficiency in the decision-making units inefficiency, representing the production technical efficiency affected by management and technology factors; *SEC* measures the scale efficiency level of decision-making units and represents the production efficiency affected by production scale factors. Equations (8) - (11) introduce the specific composition of each index.

$$EC(CRS) = \frac{D_c^{t+1}(X^{t+1}, Y^{t+1})}{D_c^t(X^t, Y^t)} \#(8)$$

$$TC(CRS) = \left[\frac{D_c^t(X^{t+1}, Y^{t+1})}{D_c^{t+1}(X^{t+1}, Y^{t+1})} * \frac{D_c^t(X^t, Y^t)}{D_c^{t+1}(X^t, Y^t)}\right]^{\frac{1}{2}} \#(9)$$

$$PEC(VRS) = \frac{D_v^t(X^{t+1}, Y^{t+1})}{D_v^t(X^t, Y^t)} \#(10)$$

$$SEC(CRS, VRS) = \frac{D_v^t(X^t, Y^t)}{D_c^t(X^t, Y^t)} * \frac{D_c^{t+1}(X^{t+1}, Y^{t+1})}{D_v^{t+1}(X^{t+1}, Y^{t+1})} \#(11)$$

where $(X^t, Y^t)$, $(X^{t+1}, Y^{t+1})$ refers to the input-output vector in period *t* and *t+1*; $D_c^t, D_c^{t+1}$ refers to the distance function based on constant returns to scale in period *t* and *t+1*, namely the compression ratio of actual output of decision making unit to optimal output; $D_v^t, D_v^{t+1}$ represents the distance function based on variable returns to scale in period $t$ and $t+1$; $D_c^t, D_c^{t+1}, D_v^t, D_v^{t+1}$ can be obtained by DEA. If $EC > 1$, the efficiency is improved; if $EC = 1$, the efficiency does not change; if $EC < 1$, the efficiency is reduced.

Based on the above, capital stock, employment and total energy consumption of sample countries are selected as input indicators. GDP is regarded as the expected output, representing the total output value; the total emission of $CO_2$ is regarded as the undesirable output, which represents a greenhouse gas emission (Li et al., 2018). The specific input-output settings are shown in Table 1.

**Table 1.** Measurement of input and output index by Energy TFP

| first grade indicators | second grade indicators | third grade indicators |
| --- | --- | --- |
| Input index | capital | Capital stock |
| | labor | Employment |
| | energy | Total energy consumption |
| Output index | desirable output | GDP |
| | undesirable output | Total CO2 emissions |

Table 2 shows the descriptive statistics of the dynamic energy TFP, and as can be seen there is huge differences in the energy efficiency and decomposition indicators when looking at the whole sample data. Further, dynamic energy TFP is calculated with the Maxdea Ultra 8.19 software. The specific results are shown in Table 3.

**Table 2.** Descriptive statistics of dynamic energy TFP

| | Item | Summary | High Income | Upper Middle Income | Lower Middle Income | Low Income |
| --- | --- | --- | --- | --- | --- | --- |
| | N | 5716 | 2046 | 1514 | 1294 | 786 |
| Dynamic TFP | Mean | 1.004 | 1.000 | 0.999 | 1.016 | 1.006 |
| | Min | 0.084 | 0.084 | 0.233 | 0.089 | 0.122 |
| | Max | 12.663 | 5.826 | 3.705 | 12.663 | 10.900 |
| (EC) Technical Efficiency Change | Mean | 1.092 | 1.108 | 1.063 | 1.092 | 1.093 |
| | Min | 0.019 | 0.043 | 0.240 | 0.092 | 0.093 |
| | Max | 16.934 | 16.934 | 3.737 | 11.491 | 10.737 |
| (TC) Technological Change | Mean | 1.006 | 1.010 | 0.985 | 0.986 | 1.008 |
| | Min | 0.060 | 0.060 | 0.246 | 0.111 | 0.218 |
| | Max | 53.576 | 18.729 | 2.859 | 3.317 | 3.626 |
| (PEC) Pure Efficiency Change | Mean | 1.036 | 1.013 | 1.040 | 1.048 | 1.036 |
| | Min | 0 | 0 | 0.027 | 0.003 | 0.204 |
| | Max | 18.707 | 11.397 | 13.837 | 13.521 | 4.081 |
| (SEC) Scale Efficiency Change | Mean | 1.082 | 1.080 | 1.076 | 1.087 | 1.075 |
| | Min | 0 | 0 | 0.072 | 0.085 | 0.120 |
| | Max | 33.435 | 16.965 | 33.435 | 32.308 | 5.595 |

As can be seen from Table 3, the average value of the dynamic energy TFP index of the whole sample countries from 1981 to 2018 is 1.004, indicating that the overall productivity is in a state of continuous growth. Among them, the energy efficiency of High-Income countries is more stable and almost unchanged; the energy efficiency of the Upper Middle-Income countries declined slightly; the energy efficiency of the Lower Middle-Income countries maintained a high growth rate for a long time, and the energy efficiency of the Low-Income countries also continued to grow. In addition, all efficiency change indexes are greater than 1 except for the Upper Middle Income and Lower Middle-Income countries with technological change index (TC) less than 1, indicating that since the 1980s, the global energy input resource allocation level and energy technology improvement are basically in a state of continuous growth.

According to the above analysis, H1 is proved in this paper. Figure 1 shows the annual trend of the mean of dynamic energy TFP and its decomposition index.

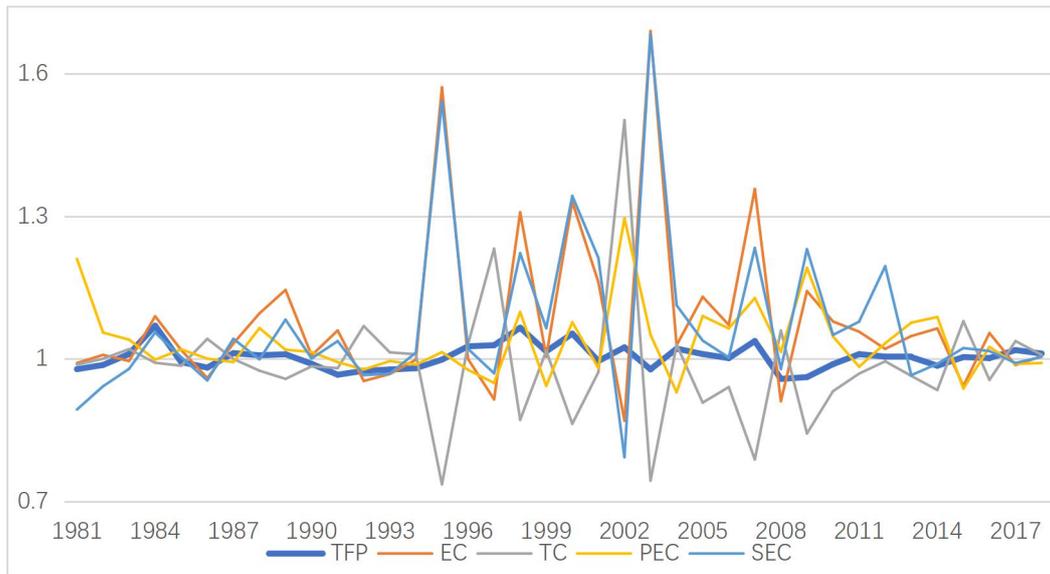

**Fig 1.** Trend line of mean value of dynamic TFP and its decomposition index

## 3.2 Descriptive statistics

In addition to the independent variable energy TFP, the other variables in this paper include input-output index, the explained variable PM2.5 concentration, the moderating variable energy investment and three control variables. Table 3 reports descriptive statistics for the remaining variables. In order to reflect the differences among different developed countries, Table 3 shows descriptive statistics of low income, lower middle income, upper middle income countries and high income countries respectively; in order to reflect the differences among different countries with different levels of science & technology development, descriptive statistics of low-tech level, medium level and high level of science & technology are displayed respectively; in order to reflect the differences among countries with different levels of new energy utilization, descriptive statistics of low proportion of new energy, medium proportion of new energy and high proportion of new energy are presented; in order to reflect the differences among countries that play different roles in international energy trade, descriptive statistics of energy importing countries and energy exporting countries are presented. The specific national classification standards of each dimension are explained in the additional analysis of the fifth section.

**Table 3.** Descriptive statistics of other variables

| | Energy-Importing countries | | Energy-Exporting countries |
|---|---|---|---|
| | N | Mean | N |
| | 4206 | 1252072 | 1434 |
| | 4206 | 10.33 | 1434 |
| | 4206 | 1.93 | 1434 |
| | 4206 | 336681.9 | 1434 |
| | 4206 | 111.38 | 1434 |
| | 1321 | 36.96 | 464 |
| | 3212 | 30324.68 | 1084 |
| | 3312 | -1551.63 | 1119 |
| | 3312 | 84.37 | 1119 |
| | 3312 | 40.67 | 1119 |

| | Lower Middle Income | | Upper Middle Income | | High Income | | Low Science & Technology | | Medium Science & Technology | | High Science & Technology | | Low New Energy Utilization | | Middle New Energy Utilization | | High New Energy Utilization | |
|---|---|---|---|---|---|---|---|---|---|---|---|---|---|---|---|---|---|---|
| | N | Mean | N | Mean | N | Mean | N | Mean | N | Mean | N | Mean | N | Mean | N | Mean | N | Mean |
| | 1294 | 1536660 | 1514 | 2690432 | 2046 | 50697.98 | 1718 | 312358.5 | 2379 | 5164333 | 1543 | 94656.86 | 1704 | 1578965 | 1942 | 2793942 | 1994 | 1665072 |
| | 1294 | 26.72 | 1514 | 8.06 | 2046 | 1.31 | 1718 | 5.68 | 2379 | 46.36 | 1543 | 1.40 | 1704 | 31.81 | 1942 | 11.60 | 1994 | 17.23 |
| | 1294 | 2.73 | 1514 | 3.73 | 2046 | 0.05 | 1718 | 0.40 | 2379 | 7.72 | 1543 | 0.22 | 1704 | 2.56 | 1942 | 3.82 | 1994 | 2.42 |
| | 1294 | 434564.8 | 1514 | 637360 | 2046 | 13111.14 | 1718 | 85292.18 | 2379 | 1302949 | 1543 | 31728.09 | 1704 | 437972.9 | 1942 | 667641.2 | 1994 | 416805.2 |
| | 1294 | 195.11 | 1514 | 211.77 | 2046 | 2.53 | 1718 | 23.70 | 2379 | 482.94 | 1543 | 12.36 | 1704 | 190.09 | 1942 | 208.48 | 1994 | 153.64 |
| | 417 | 26.29 | 511 | 20.62 | 606 | 30.57 | 490 | 31.64 | 804 | 23.35 | 491 | 35.37 | 480 | 33.39 | 630 | 20.55 | 675 | 26.29 |
| | 974 | 20371.48 | 1184 | 61458.01 | 1552 | 568.83 | 1284 | 738687.6 | 1869 | 105028.4 | 1143 | 790.94 | 1260 | 23251.66 | 1472 | 937443.8 | 1564 | 457195.9 |
| | 1006 | -3982.32 | 1217 | 1884.75 | 1602 | -125.99 | 1328 | -733.05 | 1920 | -1145.31 | 1183 | -118.45 | 1305 | -2937.85 | 1519 | 1050.10 | 1607 | -360.33 |
| | 1006 | 109.02 | 1217 | 314.63 | 1602 | 171.86 | 1328 | 140.69 | 1920 | 274.44 | 1183 | 196.54 | 1305 | 281.09 | 1519 | 86.85 | 1607 | 219.99 |
| | 1006 | 30.71 | 1217 | 28.75 | 1602 | 23.77 | 1328 | 29.69 | 1920 | 28.68 | 1183 | 26.71 | 1305 | 29.61 | 1519 | 26.55 | 1607 | 23.25 |

| Variables | Unit | Summary | | Low Income | | | |
|---|---|---|---|---|---|---|---|
| | | Mean | N | Mean | N | Mean | N |
| Capital stock | Million dollars | 1554008.7 | 5716 | 58697.36 | 786 | 712134.6 | 786 |
| Employment | Million | 15.34 | 5716 | 5.03 | 786 | 20.40 | 786 |
| Total energy consumption | Quad Btu | 2.29 | 5716 | 0.05 | 786 | 0.89 | 786 |
| GDP | Million dollars | 395893.53 | 5716 | 20071.07 | 786 | 199489.4 | 786 |
| Total CO2 emissions | MM Tonnes | 142.29 | 5716 | 2.43 | 786 | 58.22 | 786 |
| PM2.5 concentration | µg / m³ | 29.07 | 1785 | 38.03 | 251 | 39.35 | 251 |
| Energy investment | Million dollars | 345179.4 | 4352 | 7.99 | 586 | 1418766 | 586 |
| Net FDI | Million dollars | -652.63 | 4489 | -256.95 | 606 | -941.19 | 606 |
| Population density | Person / km2 | 184.63 | 4489 | 89.63 | 606 | 131.21 | 606 |
| Proportion of industrial added value in GDP | % | 27.64 | 4489 | 19.81 | 606 | 26.91 | 606 |

As can be seen from Table 3, there are significant differences in national data under different dimensions. Based on this, this paper will analyze the heterogeneity of the impact of energy efficiency on PM2.5 based on national differences in each dimension in the fifth section. In addition, to highlight the variation trend of PM2.5 concentration, the explained variable in this paper, Figure 2 shows the variation trend of the arithmetic mean of annual PM2.5 concentration and the weighted mean of GDP in the sample countries.

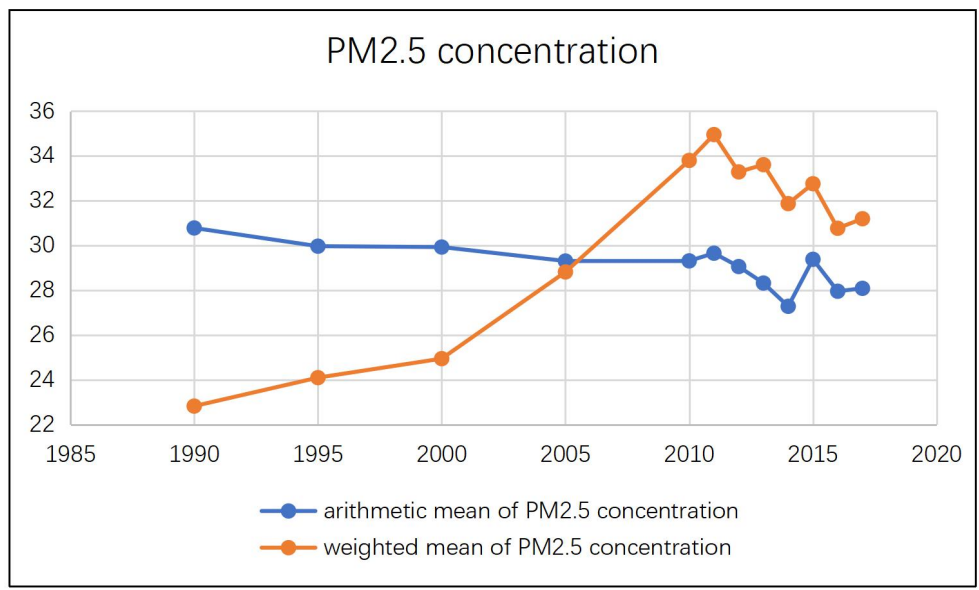

**Fig 2.** Trend line of mean value of PM2.5 concentration

## 4  Empirical results

### 4.1  Energy TFP and PM2.5

In this paper, the panel fixed effect FE model and quantile MM-QR model are used to analyze the impact of the change of dynamic energy efficiency and its decomposition indicators on PM2.5, see Equations (6)-(10). According to the model setting, the estimated results are shown in Table 4.

**Table 4.** The estimation results for the fixed effect and MM-QR models

|  |  | FE | Q10 | Q25 | Q50 | Q75 | Q90 |
|---|---|---|---|---|---|---|---|
| Dynamic TFP | (1) | -0.793 | 0.828 | 0.109 | -0.907 | -1.720* | -2.161* |
| EC | (2) | 0.295** | 0.305 | 0.301 | 0.295 | 0.290** | 0.287 |
| TC | (3) | -1.422*** | -2.813*** | -2.182*** | -1.335** | -0.602 | -0.159 |
| PEC | (4) | -0.823** | -0.697 | -0.754 | -0.834** | -0.895** | -0.928* |
| SEC | (5) | 0.086 | -0.022 | 0.025 | 0.094 | 0.146 | 0.176 |
| Controls |  | Yes | Yes | Yes | Yes | Yes | Yes |
| Country FE |  | Yes | Yes | Yes | Yes | Yes | Yes |

Notes：*, **, *** stand for significant levels of 10%, 5% and 1% respectively.

According to Table 4, the following conclusions can be drawn: the results in row (1) of table 4 shows that with TFP as the explanatory variable, the rise of dynamic energy efficiency can inhibit the emission of PM2.5 to a certain extent. In low-pollution countries, this effect is not obvious, and the inhibition effect gradually strengthens with the rise of pollution degree. In addition, combined with results (2)-(5), it can be found that this inhibiting effect mainly comes from the rise of technology level (TC) and the improvement of energy technology efficiency (PEC) brought by management and technology. According to the result of row (2) in table 4, when EC as the explanatory variable, the increase of resource allocation level of energy input promotes the emission of PM2.5, and the promoting effect decreases slowly with the increase of pollution degree. Result in row (3) from table 4, which takes TC as the explanatory variable, shows that the increase of the technological level of energy utilization inhibits the emission of PM2.5. Which means the invention of energy technology, the fundamental innovation of energy production and the improvement of technology prevent the emission of PM2.5. With the increase of the pollution level, the inhibition effect gradually decreases. As can be seen from result row (4) of table 4, with PEC as the explanatory variable, the improvement of energy technology efficiency brought about by management and technology inhibits the emission of PM2.5. With the increase of pollution level, the inhibiting effect gradually strengthens. According to results of row (5) from table 4, with SEC as the explanatory variable, the improvement of energy technology efficiency brought by the expansion of production scale does not significantly promote the emission of PM2.5, but the promoting effect increases slowly with the increase of pollution level. According to the above analysis, H2 is partially accepted in this paper.

In order to verify the robustness of the above empirical results, static energy TFP is also measured in this paper, and the relationship between energy efficiency and PM2.5 was verified by regression using the same steps. The input and output settings of static energy TFP are consistent with those of dynamic energy TFP. The DEA based on super-efficiency SBM-window is adopted, and the Window width is set to 1. In this case, the number of "DMU" in the Window is the least, equal to the actual number of DMU, which is equivalent to separating the data in each period. Then the DMU of each phase is analyzed respectively. The TFP measured by this method can only be decomposed into Pure Technical Efficiency Score (PES) and Scale Effect

Score (SES), whose meanings are similar to those of PEC and SEC mentioned above. Table 5 reports the results of the robustness test.

Table 5. Robustness test for static TFP

|  |  | FE | Q10 | Q25 | Q50 | Q75 | Q90 |
|---|---|---|---|---|---|---|---|
| Static TFP | (1) | -1.773*** | -3.418 | -2.688 | -1.685 | -0.794 | -0.257 |
| PES | (2) | -2.250*** | -3.382 | -2.905 | -2.193* | -1.564 | -1.194 |
| SES | (3) | -0.523 | -1.191 | -0.886 | -0.477 | -0.142 | 0.043 |
| Controls |  | Yes | Yes | Yes | Yes | Yes | Yes |
| Country FE |  | Yes | Yes | Yes | Yes | Yes | Yes |

Notes: *, **, *** stand for significant levels of 10%, 5% and 1% respectively.

It can be seen from Table 5 that, firstly, both static energy TFP and dynamic energy TFP can inhibit the emission of PM2.5. Secondly, the essence of static TFP is similar to the decomposition index TC of dynamic TFP. As can be seen from Table 5, the inhibition effect of static TFP and TC on PM2.5 decreases with the increase of pollution degree. Finally, the robustness test also shows that the inhibiting effect of energy efficiency on PM2.5 mainly comes from technological progress rather than scale expansion.

## 4.2 Moderating role of energy investment

Energy investment is closely related to the development of a country's energy industry. While expanding production scale, investment can also introduce or promote the progress of new energy technologies, thus affecting pollution emissions. Previous studies mainly focus on the impact of FDI on pollutant emission. Some studies believe that FDI has a positive effect on reducing PM2.5 concentration (Xie and Sun, 2020), while others believe that FDI exacerbates PM2.5 pollution, namely the "Pollution Heaven" hypothesis (Cheng et al., 2020). The main reason for the contradiction of conclusions is that FDI and PM2.5 are not directly related to each other, and the conclusions obtained from different samples and time are random and lack of universality and pertinence to pollution sources. At present, there is no unified international energy investment index standard. This paper refers to Li et al. and chooses the total assets of listed energy enterprises whose Trading Address is the target country to represent the energy investment of a country (EI) (Li et al., 2016). Table 6 reports the regression results with the decentralized product of Dynamic TFP and LnEI "C_TFP * C_LnEI" as the interaction term.

Table 6. The estimation results for moderating effect with the fixed effect and MM-QR models

|  | FE | Q10 | Q25 | Q50 | Q75 | Q90 |
|---|---|---|---|---|---|---|
| Dynamic TFP | -1.420* | -0.048 | -0.621 | -1.479* | -2.240** | -2.684* |
| LnEI | -0.156*** | -0.154* | -0.155** | -0.156*** | -0.157*** | -0.157** |
| C_TFP * C_LnEI | -0.596 *** | -0.656 | -0.631** | -0.593*** | -0.560** | -0.541 |
| Controls | Yes | Yes | Yes | Yes | Yes | Yes |
| Country FE | Yes | Yes | Yes | Yes | Yes | Yes |

Notes: *, **, *** stand for significant levels of 10%, 5% and 1% respectively.

As can be seen from Table 6, energy investment plays a significant moderating role in the impact of energy efficiency changes on PM2.5. With the increase of pollution level, the moderating effect decreases slowly. This proves that with the increase of energy investment, the inhibiting effect of dynamic energy efficiency on PM2.5 is strengthened. Using "C_EC * C_LnEI" and "C_TC * C_LnEI" as interaction terms for regression, it is found that the moderating effect of energy investment mainly plays a role in the inhibiting effect of technological progress on PM2.5. Therefore, H3 is accepted and we can claim that energy investment plays as a moderator in promoting the inhibiting effect of energy efficiency changes on PM2.5. Also, the inhibition effect of dynamic energy efficiency on PM2.5 is greatly improved with the increase of energy investment. Further, the moderating effect of energy investment decreases slowly with the increase of pollution. The moderating effect of energy investment is mainly exerted on the inhibition of PM2.5 by technological progress which also shows the importance of technological progress.

Static energy efficiency is also used to conduct a robust analysis on the moderating effect of energy investment. Table 7 reports the regression results with the decentralized product of Static TFP and LNEI "C_static TFP * C_LnEI" as the interaction term.

Table 7. Robustness test for moderating effect

|  | FE | Q10 | Q25 | Q50 | Q75 | Q90 |
|---|---|---|---|---|---|---|
| Static TFP | -1.210*** | -2.595*** | -1.970*** | -1.137** | -0.352 | 0.109 |
| LnEI | -0.094** | -0.048 | -0.069 | -0.096** | -0.122** | -0.137* |
| C_ Static TFP * C_LnEI | -0.159* | -0.191 | -0.176 | -0.157* | -0.140 | -0.129 |
| Controls | Yes | Yes | Yes | Yes | Yes | Yes |
| Country FE | Yes | Yes | Yes | Yes | Yes | Yes |

Notes：*,**,*** stand for significant levels of 10%, 5% and 1% respectively.

As can be seen from Table 7, energy investment also plays a certain moderating role in the impact of static energy efficiency on PM2.5. With the increase of pollution level, the moderating effect also decreases slowly. The conclusions are still robust.

## 5   Additional analysis

Although the results of the above quantile regression can help understand the impact of dynamic energy efficiency on PM2.5 emissions from the perspective of pollution level, pollution level alone cannot fully reflect the differences among the sample countries. In the additional analysis, this paper analyzes the heterogeneity of the impact of energy efficiency on PM2.5 from several additional dimensions (Li et al., 2020c;Li et al., 2021c;Li et al., 2021d), in order to comprehensively show the effects of exogenous factors on the impact of dynamic energy efficiency on PM2.5.

### 5.1   Heterogeneity analysis of the development level

National income level is usually the most common standard to measure the level of national development (Wang et al., 2018). According to the latest income classification of countries released by the World Bank in July 2020, this paper divides the sample countries into low-income countries, lower middle income countries, upper middle income countries, and high income countries. Simple analysis shows that the more developed a country is, the lower the PM2.5 concentration is. Fixed effect model regression is also adopted in the analysis, the model setting is the same as Equations (6)-(10). Table 8 reports the classified regression results.

Table 8. The estimation results for development level heterogeneity

|  |  | PM 2.5 | | | |
|---|---|---|---|---|---|
|  |  | Low Income | Lower Middle Income | Upper Middle Income | High Income |
| Mean of PM2.5 | | 39.03 | 38.35 | 26.29 | 20.62 |
| Dynamic TFP | (1) | 0.026 | -0.582 | 1.613 | -4.526*** |
| EC | (2) | 0.883 | 0.101 | 2.576*** | 0.183 |
| TC | (3) | -0.679 | -0.802 | -2.659*** | -1.664*** |
| PEC | (4) | -1.327 | -0.776 | 0.080 | -1.913*** |
| SEC | (5) | 1.863*** | -0.054 | 2.850*** | 0.251** |
| Controls | | Yes | Yes | Yes | Yes |
| Country FE | | Yes | Yes | Yes | Yes |
| N | | 251 | 417 | 511 | 606 |

Notes：*,**,*** stand for significant levels of 10%, 5% and 1% respectively.

As can be seen from Table 8, the impact of dynamic energy efficiency on PM2.5 varies significantly among countries with different levels of development. The more developed the country, the lower the PM2.5 concentration. According to result (1), the energy efficiency of upper middle income countries has a certain positive effect on

PM2.5, while the energy efficiency of high income countries has a stronger inhibitory effect on PM2.5. Combining results (2)-(5), it is found that although the environmental protection effect of energy efficiency in upper middle income countries is more strongly promoted by the progress of production technology, it is also negatively affected by the improvement of technical efficiency, and the negative effect mainly comes from scale efficiency. This is because the upper middle income countries, such as China, are in the period of rapid industrial development, and the scale of production is increasing rapidly, resulting pollution far outweighs the reduction in pollution caused by technological progress; In high income countries, where the industry is usually highly developed or even shifted outward, the expansion of production scale has no significant impact. At this time, the environmental protection effect of technological progress is reflected, making energy efficiency greatly restrain the emission of PM2.5. By comparison, it can be found that technical efficiency EC and technical progress TC play a decisive role in the heterogeneous impact in the dimension of country development level.

## 5.2 Heterogeneity analysis of the science & technology level

Countries with different levels of scientific & technological development tend to have heterogeneity in energy consumption due to different factors such as the advanced degree of equipment and residents' usage habits. In this paper, the average number of science & technology articles published in science and engineering journals in a country over the past years is used to represent the level of science & technology development, with data from the World Bank Database. The indicators of each country are ranked from small to large and divided into three levels: low, medium and high science & technology development. Simple analysis shows that countries with high science & technology development level have significantly lower PM2.5 concentration. Fixed effect model regression is adopted, and the form is the same as Equations (6)-(10). Table 9 reports the classified regression results.

Table 9. The estimation results for science & technology level heterogeneity

| | | PM 2.5 | | |
|---|---|---|---|---|
| | | Low Science & Technology | Medium Science & Technology | High Science & Technology |
| Mean of PM2.5 | | 30.57 | 31.64 | 23.35 |
| Dynamic TFP | (1) | -0.250 | -0.355 | -9.884*** |
| EC | (2) | 0.303 | 0.630 | 0.069 |
| TC | (3) | -0.753 | -1.103 | -1.631** |
| PEC | (4) | 0.126 | -2.262*** | -2.612*** |
| SEC | (5) | -0.033 | 1.719*** | 0.135 |
| Controls | | Yes | Yes | Yes |
| Country FE | | Yes | Yes | Yes |
| N | | 490 | 804 | 491 |

Notes: *, **, *** stand for significant levels of 10%, 5% and 1% respectively.

It can be seen from Table 9 that the impact of dynamic energy efficiency on PM2.5 is significantly different among countries with different levels of scientific & technological development. According to result (1), the environmental protection effect of energy efficiency in countries with high level of science & technology development is far greater than that in countries with medium and low level of science & technology development. Combined with results (2)-(5), it is found that the technological progress and pure technological efficiency of countries with high level of scientific & technological development are higher than those of countries with medium and low level of scientific & technological development, which is the main reason for their higher environmental protection effect of energy efficiency. In addition, countries with a high level of scientific & technological development have a small increase in national scale efficiency on pollution, mainly because the energy utilization in their production has been more mature. By comparison, it can be found that technological progress TC and pure technical efficiency PEC play a decisive role in the heterogeneous impact in the dimension of scientific & technological development level.

### 5.3 Heterogeneity analysis of the new energy utilization

Previous studies have proved that the utilization of new energy will affect the total energy use and carbon emissions to a certain extent (Zhao and Yang, 2019;Ma et al., 2021;Zhang et al., 2021). In this paper, the average annual percentage of alternative energy and nuclear energy in total energy use in a country is used to represent the degree of new energy utilization, with data from the World Bank Database. The indicators of each country are ranked from large to small and divided into three categories: low utilization ratio of new energy, medium utilization ratio of new energy and high utilization ratio of new energy. Simple analysis shows that the higher the utilization ratio of new energy, the lower the PM2.5 concentration. Fixed effect model regression is adopted, and the form is the same as Equations (6)-(10). Table 10 reports the classified regression results.

**Table 10.** The estimation results for heterogeneity of new energy utilization

| | | PM 2.5 | | |
|---|---|---|---|---|
| | | Low New Energy Utilization | Middle New Energy Utilization | High New Energy Utilization |
| Mean of PM2.5 | | 35.37 | 33.39 | 20.55 |
| Dynamic TFP | (1) | -1.588* | -0.934 | 2.192* |
| EC | (2) | 0.016 | 0.166 | 1.819*** |
| TC | (3) | 0.684 | -1.818** | -3.047*** |
| PEC | (4) | -0.298 | -3.281*** | 0.797 |
| SEC | (5) | -0.055 | 0.303 | 1.509*** |
| Controls | | Yes | Yes | Yes |
| Country FE | | Yes | Yes | Yes |
| N | | 480 | 630 | 675 |

*Notes: \*, \*\*, \*\*\* stand for significant levels of 10%, 5% and 1% respectively.*

As can be seen from Table 10, the impact of dynamic energy efficiency changes on PM2.5 is significantly different among countries with different levels of new energy utilization. According to result (1), it is unexpectedly found that energy efficiency promotes PM2.5 in countries with a high proportion of new energy utilization. Combining results (2)-(5), it is found that countries with a high proportion of new energy utilization have stronger technological progress of environmental protection effect. However, the technical efficiency of countries with a high proportion of new energy has a significant promoting effect on energy PM2.5 emissions, which comes from the promotion of pure technical efficiency and scale efficiency. On the one hand, this is because countries with a high proportion of new energy utilization have a smaller PM2.5 emission volume and are more likely to be promoted by pure energy technical efficiency and scale efficiency. On the other hand, because countries with a high proportion of new energy utilization have a relatively high and stable energy input-output ratio, it is more difficult to increase (i.e., TFP>1), so the overall fluctuation of TFP is small, and the effect of technological progress may be weaker than that of technical efficiency in reflecting its impact on PM2.5. By comparison, it can be found that technical efficiency EC plays a decisive role in the heterogeneity of the utilization degree of new energy.

### 5.4 Heterogeneity analysis of the role of international energy trade

A country's position in the global energy market also affects its own energy use to a certain extent. Based on the data of energy trade volume of sample countries from the

International Trade Centre (ITC), this paper divides sample countries into energy importing countries and energy exporting countries. Simple analysis shows that the PM2.5 concentration of energy importing countries is lower. Fixed effect model regression is adopted, and the form is the same as Equations (6)-(10). Table 11 reports the classified regression results.

Table 11. The estimation results for energy trade heterogeneity

|  |  | PM 2.5 | |
|---|---|---|---|
|  |  | Energy-Importing countries | Energy-Exporting countries |
| Mean of PM2.5 |  | 26.29 | 36.96 |
| Dynamic TFP | (1) | 0.202 | -2.084* |
| EC | (2) | 0.273** | 0.322 |
| TC | (3) | -2.058*** | 0.622 |
| PEC | (4) | -0.091 | -3.484*** |
| SEC | (5) | 0.033 | 1.679*** |
| Controls |  | Yes | Yes |
| Country FE |  | Yes | Yes |
| N |  | 1,321 | 464 |

Notes: *, **, *** stand for significant levels of 10%, 5% and 1% respectively.

As can be seen from Table 11, the impact of dynamic energy efficiency on PM2.5 is significantly different between energy importing countries and energy exporting countries. According to result (1), the change of energy efficiency of energy importing countries has no significant effect on PM2.5, while the change of energy efficiency of energy exporting countries has a significant inhibition effect on PM2.5. Combining results (2)-(5), it is found that technological progress in energy importing countries has a significant inhibiting effect on PM2.5 emissions, while technological efficiency has a more significant promoting effect on PM2.5 emissions. However, pure technical efficiency of energy exporting countries inhibited PM2.5 more obviously, while scale efficiency promoted PM2.5 emission. By comparison, it can be found that pure technical efficiency change PEC plays a decisive role in the heterogeneity of energy importing and exporting countries.

According to the above analysis results, H4 is accepted in this paper.

## 6 Conclusions and implications

First of all, global energy efficiency continues to increase during the sample period from 1980 to 2018. First, through the analysis of dynamic energy TFP, it is found that the overall energy efficiency increased by 0.4% per year on average. Second, through the analysis of Technical Efficiency Change, it is found that the energy technical efficiency increased by 9.02% on average every year, while the decomposition indicators of pure technical efficiency and scale efficiency increase by 3.6% and 8.2% on average every year respectively. Third, through the analysis of Technological Change, it is found that energy technological progress increased by 0.6% on average every year.

Secondly, the improvement of energy efficiency will inhibit PM2.5 pollution. First, there is a negative relationship between dynamic energy efficiency and PM2.5 concentration, and the same conclusion can be obtained from static energy efficiency. Second, the higher the pollution level, the greater the inhibition of PM2.5 concentration by energy efficiency, that is, the stronger the environmental protection effect. Third, the inhibition of energy efficiency on PM2.5 concentration mainly comes from technological progress and the improvement of pure technical efficiency, while the overall technical efficiency will promote the emission of PM2.5.

Thirdly, energy investment plays a moderating role in strengthening the inhibition effect of energy efficiency on PM2.5 concentration. First, energy investment plays a significant moderating role in the inhibition effect of rising energy efficiency on PM2.5. With the increase of energy investment, the inhibition effect of dynamic energy efficiency on PM2.5 is greatly improved. Second, with the increase of pollution, the moderating effect of energy investment decreases slowly. Third, the moderating effect of energy investment is mainly exerted on the inhibition of PM2.5 by technological progress.

Finally, the impact of energy efficiency on PM2.5 concentration is heterogeneous, which is reflected in the differences of national attributes. First, in the comparative analysis of the development level, the energy efficiency of upper middle income countries has a certain positive effect on PM2.5, while the energy efficiency of high income countries has a stronger inhibition effect on PM2.5. This is because although the environmental protection effect of energy efficiency of upper middle income countries is more enhanced by the progress of production technology, it is negatively affected by the improvement in technological efficiency, which the high-income countries have almost no effect on. In this dimension, technical efficiency change EC and technological change TC jointly play a decisive role. Second, in the comparative analysis from the dimension of the level of scientific & technological development, the environmental protection effect of energy efficiency of countries with high level of scientific & technological development is far greater than that of countries with medium and low level of scientific & technological development, mainly because they have higher technological progress and pure technological efficiency. In this dimension, technical progress TC and pure technical efficiency PEC play a decisive role. Third, in the comparative analysis from the dimension of the utilization ratio of new energy, the energy efficiency of countries with a high proportion of new energy utilization plays a positive role in promoting PM2.5. This is because although they have technological progress with stronger environmental protection effect, they still have pure technical efficiency and scale efficiency that significantly promote PM2.5. In this dimension, technical efficiency change EC plays a decisive role. Fourth, in the comparative analysis from the dimension of energy trade, the energy efficiency of energy exporting countries has a stronger environmental protection effect. This is because although the technological progress of energy importing countries has a greater inhibitory effect on PM2.5, the technical efficiency promotes PM2.5 emission more significantly, while the pure technical efficiency of energy exporting countries has a stronger environmental protection effect. In this dimension, pure technical efficiency PEC plays a decisive role.

The conclusion of this paper can be used as a reference for energy and environmental protection policy making in various economies. First of all, the analysis of this paper shows that the improvement of energy efficiency has basically played a positive role in the environmental protection of the economy, so countries in energy consumption and daily production should pay attention to both technical progress and technical efficiency. They should not only control the input-output level of energy allocation, but also vigorously develop and improve the energy technologies that are being used. In the future, disruptive new technologies are likely to appear in oil and gas, hydrogen energy, energy storage, nuclear fusion energy and other fields, which will fundamentally change the picture of future energy pollution. It is of great significance to accurately grasp the development trend of energy technology for guiding the direction of scientific and technological innovation, as well as for the country to formulate energy and environmental policies and enterprise green strategic transformation. At the same time, the level of investment in energy should also be increased appropriately. Secondly, this paper empirically found that the environmental protection effect of energy efficiency is more obvious in countries with high pollution levels, which reflects the catch-up effect in the tail, and also sounds the alarm to countries with low pollution levels. The additional analysis of this paper also clearly reflects the heterogeneity of national energy efficiency and environmental effects in each dimension, and gives the decisive influencing factors based on differences. Therefore, countries should make strategic adjustments to their energy and environmental protection policies according to their own conditions (Kawabata, 2020;Liu et al., 2020). For example, countries with low development level need to improve energy efficiency in all aspects, while countries with high proportion of new energy utilization should focus on improving their energy input-output resource allocation level. In general, countries need to start from these perspectives, according to their situation, implement internal and external policies of energy, environmental protection, science & technology and the supply side to improve energy efficiency and environmental protection effect step by step (Li et al., 2021b), steadily improve their own pollution in the long term, and spare no effort to maintain the earth's environment and public health.

**Conflict of Interest**

The authors declare that the research was conducted in the absence of any commercial or financial relationships that could be construed as a potential conflict of interest.

**Funding**

This research was funded by the 13th Five-year Plan fund for the development of Philosophy and Social Sciences of Guangzhou in 2020 "Research on statistical measurement and evaluation of innovation factor allocation from the perspective of high-quality development", grant number 2020GZYB89.